# Role of Isospin Degree of freedom on the impact parameter dependence of balance energy

*Dolly Sood[1], Sanjeev Kumar[2], Suneel Kumar[2] and Rajeev K. Puri[3]

[1]Shree Ganesh Group of institute, Rakhra, Patiala, Punjab, India.
[2]School of Physics and material science, Thapar University, Patiala 147004, Punjab, India.
[3]Department of Physics, Punjab University, Chandigarh, India.
* email:dolly.sood19@gmail.com

## Introduction

The study of heavy ion collisions is a rapidly expanding subject. Heavy ion physics has attracted much attention during last three decades. A large number of accelerators have been developed to study these heavy ion reactions. Heavy ion collisions (HIC) at intermediate and high energies provide a possibility for studying the properties of nuclear matter in conditions vastly different from that in normal nuclei, such as high density, high temperature and excitation as well as large differences in proton neutron numbers. The nuclear equation of state (EOS) can be researched via liquid gas transition, multifragmentation and collective flow, such as directed and elliptic flow produced in heavy ion collisions. Such knowledge is not only of interest in nuclear physics but is also useful in understanding astrological phenomenon such as evolution of early universe. One observable that has extensively used for extracting nuclear (EOS) from heavy ion collisions at intermediate and high energies is the collective flow. Due to large difference in the proton neutron number, isospin degree of freedom comes into account in intermediate energy heavy ion collisions. The isospin degree of freedom is supposed to play a very important role on the collective flow. For the determination of the equation of state (EOS), the relationship between pressure and volume for nuclear matter, is an important objective of nuclear physics. The information about the equation of state can be extracted from the collective flow of nuclear matter deflected sidewards from the hot and dense region formed by the overlap of projectile and target nuclei [1].

## The Model: IQMD

The present study is carried out within the framework of Isospin dependent Quantum Molecular Dynamical (IQMD) model [2] also known as Semi classical microscopic improved version of QMD model [3] which is based on event by event method & includes symmetry potential, isospin dependent NN cross-section. Heavy–ion collisions are simulated by generating the phase space (x, y, z, $p_x$, $p_y$, $p_z$) of two colliding nuclei at different time steps.
Three steps of simulation are:

1. **Initialization of nucleus:** In this step, each nucleon is represented by Gaussian wave packet given as

$$f_i(\vec{r},\vec{p},t) = \frac{1}{(\pi\hbar)^3} \times e^{[-(\vec{r}-\vec{r}_i(t))^2 \frac{2}{L}]} \times e^{[-(\vec{p}-\vec{p}_i(t))^2 \frac{L}{2\hbar^2}]}$$

2. **Propagation of ($A_T + A_P$) nucleon system:**

Each nucleon propagates under the classical Hamiltonian's equations of motion, given by

$$\frac{dr_i}{dt} = \frac{\partial \langle H \rangle}{\partial P_i}, \qquad \frac{dP_i}{dt} = -\frac{\partial \langle H \rangle}{\partial r_i}$$

Where $\langle H \rangle = \langle T \rangle + \langle V \rangle$ is Hamiltonian.

The total interaction potential is given as:
$$V_{ij} = V^{ij}_{Skyrme} + V^{ij}_{Yuk} + V^{ij}_{Coul} + V^{ij}_{mdi} + V^{ij}_{sym}$$

3. **NN collision:** Collision between two nucleon takes place if the following condition is satisfied.

$$d < |r_i - r_j| = \sqrt{\frac{\sigma_{nn}}{\pi}}$$

## Results and discussion:

For this study we have simulated the $_{28}Ni^{58}+_{28}Ni^{58}$ and $_{26}Fe^{58}+_{26}Fe^{58}$ systems at incident energies ranging from 45 MeV/nucleon to 105 MeV/nucleon by using soft equation of state. The geometry of reactions chosen are $\hat{b} = b/b_{max}$ = 0.2, 0.4, 0.5, 0.6, and 0.7.

In fig. 1&2, average $<p_x/A>$ as a function of scaled rapidity $y_{c.m}/y_{beam}$ is shown. The results are displayed at impact parameters $\hat{b}$ = 0.2, 0.4, 0.5, 0.6 and 0.7. In all the cases, + ve to – ve transition is observed for all impact–parameters upto 85Mev/nucleon. However at higher energies, -ve to +ve transition is observed for $\hat{b}$ =0.2 (central collisions). This is indicated that the balance energy for $\hat{b}$ =0.2 is between 85 to 95 Mev/nucleon. Moreover with increase in impact-parameter, the slope is increasing at all incident energies. From here it is clear that balance energy is sensitive towards the impact-parameter or geometry of reaction [1].

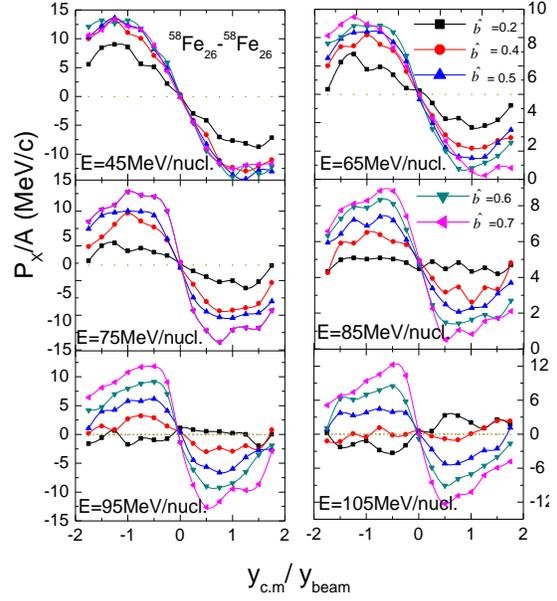

Fig 2. same as fig.1 but for $_{26}Fi^{58}+_{26}Fi^{58}$ system.

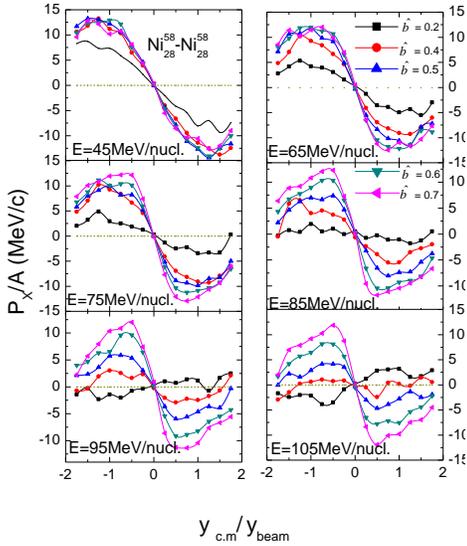

Fig 1. The averaged $P_x/A$ as a function of $Y_{c.m}/Y_{beam}$ with σ=80% of free cross section.

If one compares the fig. 1 and fig.2 the shape of lines are similar in both the figures. The difference observed is in the magnitude of $P_x/A$. The $P_x/A$ is becoming more positive for neutron – rich system $_{26}Fi^{58}$ as compared to neutron-poor system $_{28}Ni^{58}$. This point towards the higher value of balance energy for $_{26}Fi^{58}$ system. Further study in this direction is ongoing.

**References:**
[1] R.Pak et al., phys. Rev. Lett. **78** 1026 (1998).
[2] C. Hartnack, R. K. Puri, J. Aichelin, J. Konopka, S. A. Bass, H. Stocker, W. Greiner, Eur. Phys. J A 1 **151** 169(1998).
[3] J. Aichelin, Phys. Rep. 202, 233 (1991).